\documentclass[]{interact}

\usepackage{epstopdf}
\usepackage[caption=false]{subfig}

\usepackage[natbibapa,nodoi]{apacite}
\usepackage{cleveref}

\theoremstyle{plain}

\theoremstyle{definition}

\theoremstyle{remark}

\begin{document}

\articletype{}

\title{Towards robust and speculation-reduction real estate pricing models based on a data-driven strategy}

\author{
\name{Vladimir Vargas-Calderón\textsuperscript{a, b}\thanks{CONTACT Vladimir Vargas-Calderón. Email:vvargasc@unal.edu.co} and Jorge E. Camargo\textsuperscript{c}}
\affil{\textsuperscript{a}Laboratorios de Investigación en Inteligencia Artificial y Computación de Alto Desempeño, Human Brain Technologies, Bogotá, Colombia;\\ \textsuperscript{b}Grupo de Superconductividad y Nanotecnología, Departamento de Física, Universidad Nacional de Colombia, Bogotá, Colombia;\\ \textsuperscript{c}System~Engineering~Department, Fundación~Universitaria Konrad~Lorenz, Bogotá, Colombia}}

\maketitle

\begin{abstract}
In many countries, real estate appraisal is based on conventional methods that rely on appraisers' abilities to collect data, interpret it and model the price of a real estate property. With the increasing use of real estate online platforms and the large amount of information found therein, there exists the possibility of overcoming many drawbacks of conventional pricing models such as subjectivity, cost, unfairness, among others. In this paper we propose a data-driven real estate pricing model based on machine learning methods to estimate prices reducing human bias. We test the model with 178,865 flats listings from Bogotá, collected from 2016 to 2020. Results show that the proposed state-of-the-art model is robust and accurate in estimating real estate prices. This case study serves as an incentive for local governments from developing countries to discuss and build real estate pricing models based on large data sets that increases fairness for all the real estate market stakeholders and reduces price speculation.
\end{abstract}


\section{Introduction}

Many countries have legally defined standard appraisal or valuation methods for the housing industry~\citep{adair1987direct,trojanek2010methodic,chau1995valuation,wong1998evolution,walacik2013valuation,abidoye2016survey,mcparland2002valuation,schnaidt2012german,dos2002practice,cushman1940judicial}. The legal establishment of such methods is due to two main reasons. The first one is because governments use the valuation for taxation purposes \citep{Case1978,cochland1874,moore2009}. The second one comes from the need to protect real estate buyers and sellers from frauds, which may compromise private investments as well as public resources. For instance, a seller may not know the market value of his/her real estate, and a buyer may take advantage of this to buy it way below the market value. Even though the seller freely accepted the buyer's offer, the seller did so with no knowledge of the market conditions. Another common situation occurs when a person tries to get a loan from a bank~\citep{bucknall2008real,cole2002automated,cagan2009method,walker2013method,carswell2009effects}. A usual loan policy is to lend a fixed percentage of the real estate market value, which is estimated through one of the admitted appraisal methods. The person interested in buying a real estate property may reach an agreement with the seller for a certain amount of money but can fix the appraisal to show a higher market value, looking for the bank to lend him/her more money.  Therefore, having an appraisal method can show the seller and the buyer an estimate of the property market value. However, such a method has to be standardised to avoid further frauds. Because of situations like these (there is a vast amount of examples), appraisal documents are legally binding documents, normally signed by a professional in the appraisal techniques, who assumes legal responsibility for the performed appraisal.

Even though such legal responsibility discourages appraisal professionals from any corruption-related activity, the commonly accepted appraisal methods (comparison, profits and residual methods) rely on geographically local market studies that strongly depend on the information that the appraisal professional is able to collect, as well as on his/her criteria to exclude some of that information in the market analysis~\citep{kucharska2013uncertainty}. Therefore, it could be the case that the honest work of two appraisal professionals lead to different market values for the same real estate property~\citep{brzezicka2014identifying,kucharska2014reproduction,mohammad2018valuer,PREVEDEN2015ExpertiseAD}, especially in developing countries where real estate databases are not public. This potential error can be mitigated in cases where enough data is available to build robust mathematical pricing models for the real estate market. With a mathematical model like the proposed in this paper, which ensures the inclusion of all the relevant data that allows a more precise and standardised appraisal of a real estate property, even when the available data is not official. Such is the case of many developing countries, where no public databases of real estate transactions exist. Nonetheless, in these countries a large amount of information can be found in real estate websites where sellers offer their properties.

It is only natural to apply data mining techniques to the available information to extract meaningful features that can be used by machine learning algorithms to estimate the market value of real estate properties correctly. In fact, many companies in developed countries claim to use artificial intelligence tools to appraise real estate properties~\citep{humphries2014automatically,rappaport2010method,abileah2014providing}, but these tools are not published for the benefit of the people. Other academical efforts have suggested mathematical models with good results, but limit their analysis to an overall view of the model, rather than examining how the model performs at different appraisal tasks, which indicate the model's quality.

It is essential that real estate market stakeholders, which include buyers, sellers, moneylenders and governments, discuss and develop appraisal tools that ensure a fair market for everyone. This is particularly relevant for cities where there exist few and vague regulations of real estate prices. Such is the case of Colombian cities, where only during the past decade an open registry of appraisal professionals has been built in order to give legal validity to appraisal reports.

In view of the lack of real estate pricing models trained with large databases in Colombia, and in general, in developing countries, we present a machine learning model based on XGBoost~\citep{Chen:2016:XST:2939672.2939785} tuned with Optuna~\citep{optuna_2019}, trained with data collected from  178,865 flats in Bogotá published in real estate websites from 2016 to 2020. The contribution of our work is two-folded. First, we apply state of the art techniques to build a machine learning model for real estate prediction. Second, we minutely describe the behaviour of the prediction taking into account geographical, price and socioeconomic stratum variables, allowing a robust assessment of the model. Therefore, our model becomes the most advanced appraisal model in Colombia, as far as Google search shows.  We hope that this paper sets a milestone that lures local governments from Colombian cities (and cities alike in developing countries) to transition from expensive, inaccurate and subjective ways of doing real estate appraisal towards the construction of data-driven alternatives that are more robust and fairer.

The paper is divided as follows. \Cref{sec:review} reviews the main works on machine learning applied to the real estate market. In \cref{sec:method}, the method is outlined, and the collected data is presented. \Cref{sec:results} presents the results and discusses them. Finally, the work is concluded in \cref{sec:conclusions}.

\section{Literature Review\label{sec:review}}

In this section, we describe the commonly used comparison method, which is typically used when appraising flats in cities because the availability of data allows a straightforward market analysis. We also mention some efforts to apply machine learning techniques to the appraisal of real estate.

\subsection{Comparison Method}

Assessing the market value of merchandise can be done by asking the suppliers for the market value of the merchandise they sell. Competition between suppliers of the same merchandise makes the sale prices distribution to be thin. However, when the merchandise is a real estate property, this scenario changes completely. The most noticeable change, which triggers many others, is that real estate properties are unique. Even when the architecture of two properties is the same, their states can be different. Real estate properties are inherently variable and unique, which is why assessing the market value of a property is a difficult task. From now on, market value will be used indistinguishably from price.

However, a method that can lead to an estimation of the price of a target property is to compare it with other properties of known price, which are as similar as possible to the target property. One can look for properties with similar features such as the number of bathrooms, the number of rooms, the area, the material of the floors within the property, etc. It is up to the appraisal method designer to choose which features to take into account and how to measure the difference between two real estate properties based on those features. Formally, the appraisal method defines a distance $d(a,b)$ (e.g. see the work by \citet{mccluskey1997evaluation}) between all possible properties $a$ and $b$, then measures the distance from the target property $t$ to other properties, and then decides to keep those properties $p_n$ for which $d(t, p_n) < \text{threshold}$, for some selected threshold. Therefore, one ends up with a set of properties $\{p_n\}$, and the price of the target property is defined under some rule function as $f(\{p_n\})$. A widely used function by appraisal professionals in Colombia is to average the price per square meter of the properties $\{p_n\}$, but more complex ones can certainly be used. The interested reader is referred to the work by \citet{pagourtzi2003real} for an example of a complex rule function.

This method works well when there is enough data to draw statistics out of it. The more data, the more rigorous the method application is. However, it is a common practice to discard properties if their prices describe a two-peaked distribution, or their distribution has long tails. In the end, it is also common to see appraisal reports where the price of a property is estimated using 5-10 properties, which is far from being a sample with stable statistics. In fact, there have been proposals to improve the sampling of similar properties~\citep{adair1987direct,vandell1991optimal,gau1992optimal}, but none have been standardised throughout the world.

On the other hand, cities are rich in data nowadays, and even in developing countries, there are real estate websites with historical information about market values that can be retrieved. If those data sources were used responsibly, probably the appraisals by the comparison method would achieve better quality. Moreover, the possibility of building a database with tens of thousands or even hundreds of thousands of real estate properties gives entrance to data mining techniques such as machine learning models that exploit the complexity and variety of data.

\subsection{Machine learning models}

\citet{trawinksi2017} have shown that machine learning models can significantly outperform expert systems such as the comparison method, even though a tuning of the machine learning models hyperparameters is not performed. In fact, researchers have for long identified the capability of machine learning models to take advantage of large amounts of data to derive regression rules that enable them to predict housing prices accurately~\citep{pagourtzi2003real}. Although in the early application of artificial neural network (ANN) models there was a debate of whether or not these could outperform hedonic or expert system models~\citep{worzala1995exploration,borst1991artificial,do1992neural,evans1992artificial,KATHMANN1993373}, it is now clear due to computational power and data availability, robust machine learning models can easily outperform hedonic and expert models~\citep{peterson2009neural,Kok202,limsombunchai2004house,trawinski2017comparison}.

In order to determine the best of many possible machine learning models for real estate appraisal, comparison studies have been performed. For instance, ~\citet{graczyk2010} implement ANNs, decision trees (DTs), linear regression (LR), and support vector machines (SVMs), concluding that no single algorithm produces the best results, though it must be mentioned that a data set of only over 1,000 transactions was used. Also, \citet{PARK20152928} analysed algorithms such as C4.5 DT, RIPPER, Naïve Bayes regression and AdaBoost, but the error rate does not allow to conclude which algorithm is the best. Here, also a small data set is used (over 5,000 transactions). Several machine learning algorithms are studied by \citet{lasota2010comparison} in a data set of over 5,000 transactions, establishing that evolutionary fuzzy rule learning performed worse than SVMs, ANNs and DTs, and later on they also concluded that random forest could outperform random subspace~\citep{lasota2011investigation}. Similarly, \citet{borde2017real} examined LR, K nearest neighbour regression and random forest regression on a data set of over 5,000 properties scrapped from Indian real estate websites for prediction of real estate prices, where the random forest outperformed the rest of the algorithms. All in all, even though these comparative studies over small data sets do not extensively perform hyperparameter optimisation, it was constant that ANNs and random forest always performed well. 

More advanced works used these techniques (ANNs ~\citep{LIU2011s626,pagourtzi2007real,CHIARAZZO2014810,din2001,cechin2000real,SELIM20092843,khalafallah2008,antonios2019,rahman2019artificial} and random forests~\citep{adgeo-45-377-2018,ceh2018}) to build housing price models, although SVMs have also been exploited~\citep{WANG20141439}. Among the most interesting works using ANNs, the following highlight. \citet{guo2018homogeneous} proposed methods for homogeneous feature transfer and heterogeneous location fine-tuning to be able to compare properties from different cities, enriching the data that can be collected by having multiple sources. \citet{bin2017regression} outline a method that uses boosting trees to predict the price of a real estate property, while a Long Short-Term Memory neural network is used to extrapolate the result into the future by estimating future housing price indices. Even Bayesian learning has been applied to ANNs to improve the performance of real estate pricing models \citep{giudice2017}. Also, convolutional-based neural networks have been used \citep{paio2019}. \citet{rafiei2016novel} used deep belief restricted Boltzmann machines to assess whether a company should build or not housing, depending on the predicted housing price. Despite the numerous works done with ANNs, lately, tree ensemble models are overtaking ANNs as the best models to estimate real estate market values. For instance, since its first application, \citet{ANTIPOV20121772} showed the superiority of random forests over multiple regression analysis, ANNs, boosted trees, K nearest neighbours, among others. \citet{peng2019} comparatively studied multiple linear regression, decision trees and XGBoost models, finding the later to be the most robust and accurate one. A very complete work was presented by \citet{denadai2018}, where an XGBoost model is trained with multi-modal data from several sources like OpenStreetMap, Google Street Views, property taxes data, census data and home listings, reaching outstanding results because of the combination of several sources of information that also have been used, in isolation, like images~\citep{liu2018learning,you2017image,poursaeed2018vision,zhao2018deep}, text~\citep{vargas2019model,pfeifer2019} and land use information~\citep{fu2015}. Different kind of studies are those based on ensemble models, which normally reach slightly better results~\citep{xiong2020,niu2019}. 

In Colombia, open research in this area is precarious, and only a couple of works exist. \citet{perez2019machine} proposed an expert system based on regression trees and linear regression algorithms (hedonic pricing models) in two large data sets of around 60,000 properties, including properties from Colombia. Their expert system can reach similar results as the hedonic pricing models taking into account the most significant variables. However, by using regression trees and not random forests, the study falls short on implementing state of the art machine learning techniques, which may have improved the results significantly. A much more advanced work is the one proposed by~\citet{martinez2019approach}, where a comparative study between decision trees, random forests and gradient boosting is done with a small data set of properties in Cali, Colombia.

Therefore, it can be concluded that machine learning models are better than expert systems. Even though ANNs are extensively used as housing price models, random forest-like models have produced state of the art results, motivating the use of XGBoost in our study.

\section{Method and Materials\label{sec:method}}
In this section, we describe the collected data and also outline the machine learning method used for building a real estate price model.
\subsection{Data}
We have collected data from one of the most popular real estate websites in Colombia from 2016 to 2020, regarding information about properties for sale in all of the Colombian territory. In this work, we focus on the city of Bogotá, and we place our study in the submarket of flats, as it is the largest and most diverse one. For each flat, we have (apart from some missing values) its price, area, number of lifts, number of bedrooms, number of bathrooms, number of closets, number of parking spots, publishing date, social class or stratum (the Colombian government classifies each property in one of six different social classes for taxation purposes), existence of private security, existence of a laundry room, location (latitude and longitude), floor number, age, type of floor of the bedrooms, type of floor of the majority of the flat, existence of water heating system, type of kitchen gas and existence of an independent dining room.

\begin{figure}
    \centering
    \includegraphics[width=\textwidth]{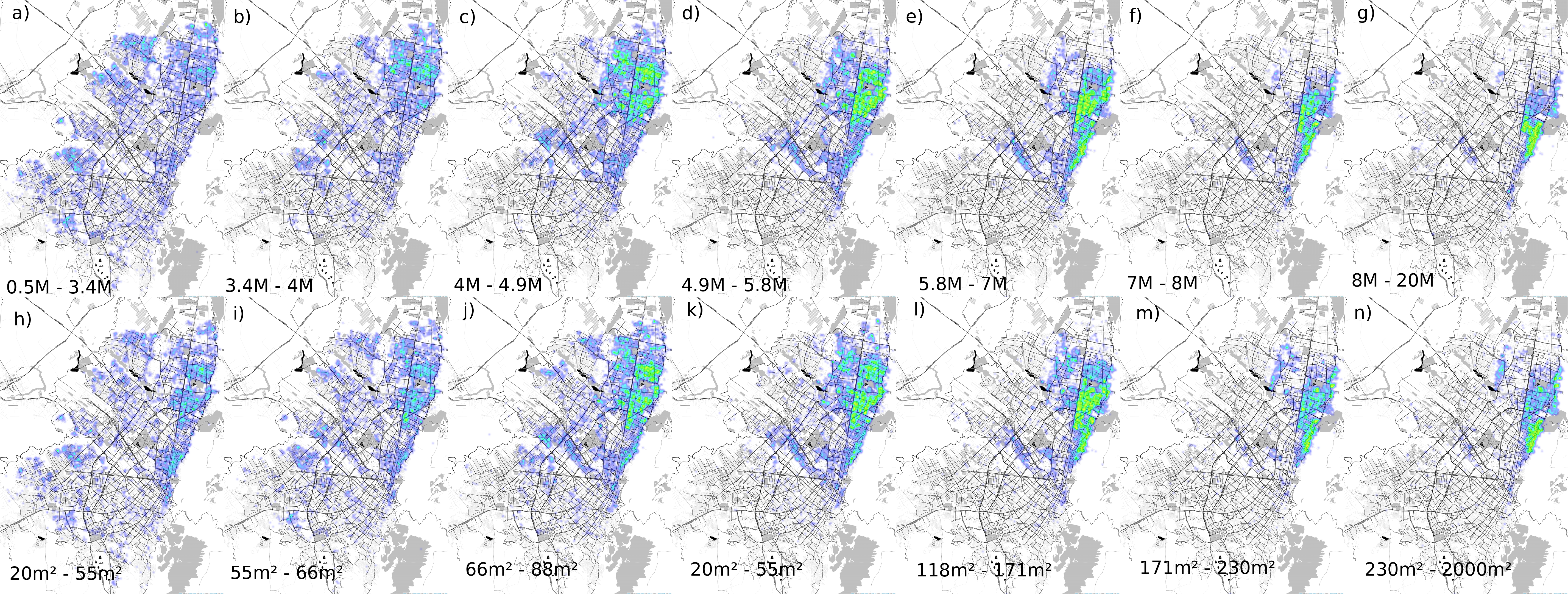}
    \caption{Heat maps of flats' geographical distribution in Bogotá. Panels a)-g) show the distribution of flats in a given price per square meter region corresponding to the 10th, 20th, 40th, 60th, 80th, 90th and 100th percentile. Each panel shows flats whose price per square meter lies within a given range of millions of Colombian pesos (COP). The situation is similar for panels h)-n), where flats are divided by construction area in the same percentiles, and ranges of areas are shown.}
    \label{fig:flatsinbogota}
\end{figure}

A total of 178,865 flats compose our database. \Cref{fig:flatsinbogota} show the geographical distribution of flats partitioned by area and price, showing that the largest and more expensive flats are found at the eastern part of the city, which is limited by mountains. In that part of the city, businessmen and women live. It is also seen that flats affordable to the middle class are located in the northeast part of the city, whereas the working class (in all its spectrum) tend to live in the west, south, and some parts of the north.

It is worth mentioning that the distribution of flats for sale in \cref{fig:flatsinbogota} refers only to those published at the studied real estate online platform. In reality, the market of the cheapest flats is not very present in these platforms, since their owners prefer to attract clients with the sale sign on the flat's windows, instead of posting them online.

\begin{figure}[!ht]
    \centering
    \includegraphics[width=0.75\textwidth]{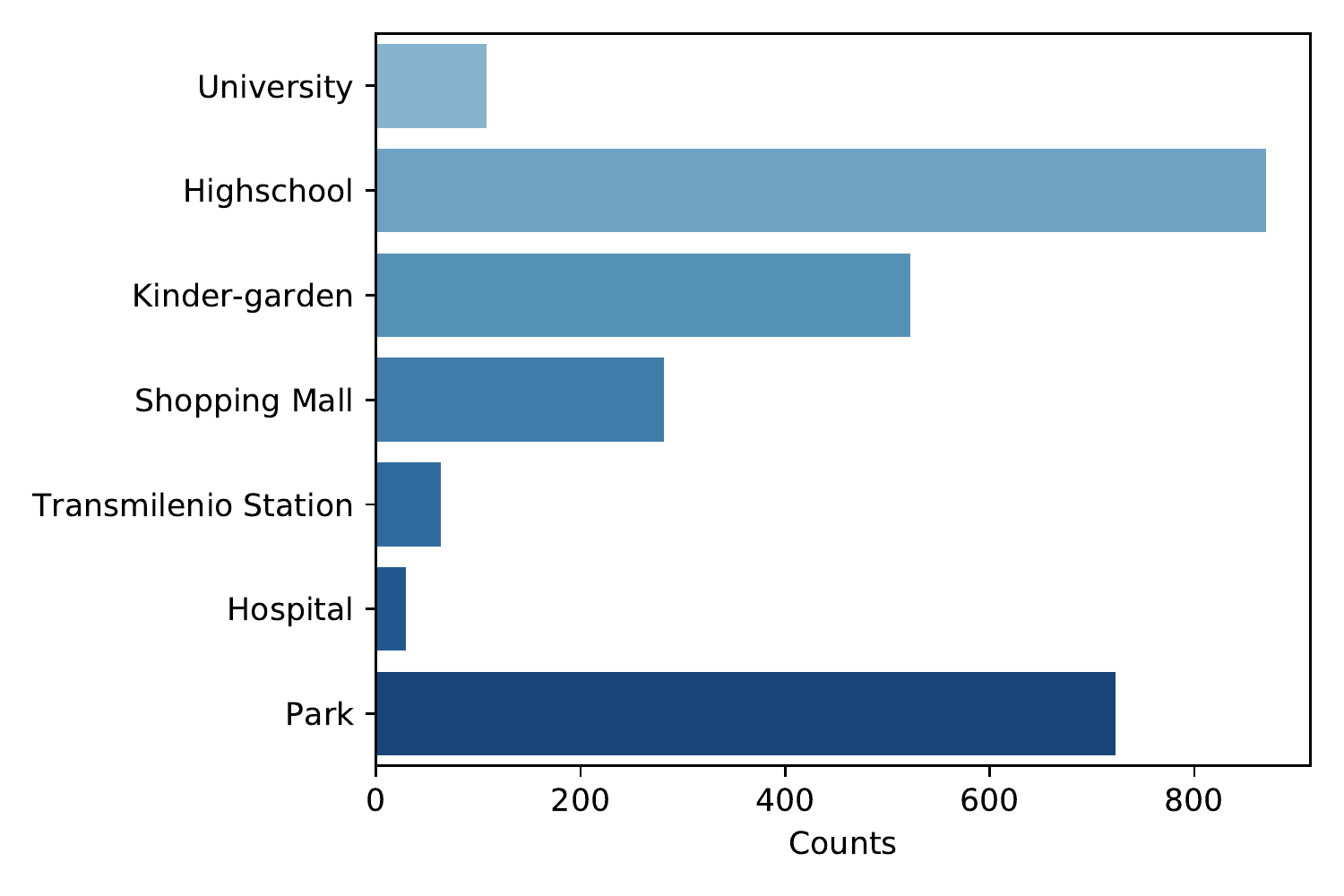}
    \caption{Number of public places found in Bogotá for each category.}
    \label{fig:count_public_places}
\end{figure}

The geographical structure of flats that correspond to different regions of price and area shows a strong spatial correlation, which can be further exploited by making use of other available geographical data such as the location of public places including shopping malls, hospitals, schools, universities, parks, public bus stops and kinder-gardens. We downloaded information from the Google Places API about these public sites in order to expand the set of features that describe every flat. The number of sites found per category is shown in \cref{fig:count_public_places}. Every flat is now described by its features plus its distance to the nearest public places of each category.

\subsection{Method}

At first, a preprocessing stage is performed. This stage consists of standardising the range of values that continuous variables take, and of encoding categorical variables to one-hot encoded vectors. The standardisation is such that the mean of the data on each numerical data is taken to be 0, and its standard deviation to be 1.

After that, we use XGBoost as a regression model to predict the price per square meter of the flats. XGBoost is a gradient-boosted algorithm based on classification and regression trees (CARTs). The main idea of XGBoost consists of building a family of weak regressors $\{f_k\}$, where $F(\vec{x})=\sum_k f_k(\vec{x})\mapsto \hat{y}$ is a function that maps an input vector of features $\vec{x}$ to a predicted value $\hat{y}$. The functions $f_k$ that compose $F$ are parameterised by a set of parameters that we denote by $\vec{\theta}_k$, and these are optimised to minimise a certain loss function. In the case of a dataset of pairs $\{\vec{x}_i, y_i\}$, where $\vec{x}_i$ are feature vectors and $y_i$ are target values, the loss function can be written as
\begin{align}
    E(\vec{\theta}) = \sum_i ||\hat{y}_i - y_i|| + \sum_k \Omega(f_k), 
\end{align}
where $||\cdot||$ is some norm and $\Omega(f_k)$ is a penalising function for the complexity of the corresponding tree represented by $f_k$ that assumes the form $\gamma T + \lambda/2 \sum_{j=1}^T w_j^2$, where $\gamma$ and $\lambda$ are regularisation parameters, $T$ is the number of leaves of the corresponding tree, and $w_j$ is a score assigned to each leave onto which the data are projected. Therefore the penalising function keeps the leaf scores small, and tries to keep the number of leaves small as well. Details of the so-called additive training can be found in Refs.\citep{Chen:2016:XST:2939672.2939785, friedman2000}.

The XGBoost model was trained under a 10-fold cross-validation framework, where a total of 30,000 flats were randomly picked as the test set.  Hyperparameter optimisation was performed with Optuna~\citep{optuna_2019}, taking the mean squared error of the price as the objective function to minimise. A total of 200 points in the parameter space were sampled.

Additionally, we assume that the model does not perform with the same quality for different types of flats. Therefore, for each price estimation of the XGBoost regressor, a search is performed in the training database, where a total of 20 flats are retrieved. These 20 flats have the closest Euclidean distance in the features space to the flat that is being studied. Therefore, the 20 flats are not necessarily close geographically, but they are in a mathematical neighbourhood around the target flat. It is expected that the XGBoost regressor performance in this neighbourhood is similar to its performance at the point where the target flat lies in the feature space. Therefore, an estimation of the error on the price estimation for the target flat can be done by measuring the error on the price estimation for the flats within the mathematical neighbourhood. Let $\{\vec{x}_n\}$ be the set of 20 flats nearest to the target flat $\vec{x}_t$, then, the error of the price per square meter on a flat from the neighbourhood is
\begin{align}
    E_n = \hat{y}_n-y_n.\label{eq:error}
\end{align}
We define the price estimation of a given flat represented in the feature space as $\vec{x}_t$ as a tuple $(p_\ell,p,p_u)$, where $p$ is the price estimated by the XGBoost regressor, and $p_{\ell(u)}$ is the lower(upper) price bound. These are computed as follows. Let $d(\vec{x}_t,\vec{x}_n)$ be the distance between the target flat and the $n$-th flat from the neighbourhood of the target flat. Then the lower price bound $p_\ell$ is
\begin{align}
    p_\ell = p + \sum_{n_\ell} E_{n_\ell} w_{n_\ell},
\end{align}
where $n_\ell$ indexes the flats from the neighbourhood of the target flat with negative error (as in \cref{eq:error}), and $w_{n_\ell}$ is defined as
\begin{align}
    w_{n_\ell} = \left. d^{-2}(\vec{x}_t,\vec{x}_{n_\ell}) \middle/ \sum_{n'_\ell} d^{-2}(\vec{x}_t,\vec{x}_{n'_\ell}). \right.
\end{align}
Similarly, the upper price bound  is defined as $p_u = p + \sum_{n_u} E_{n_u} w_{n_u}$, where $n_u$ indexes the flats from the neighbourhood of the target flat with positive error.

\section{Results and Discussion\label{sec:results}}

\begin{figure}[!ht]
    \centering
    \includegraphics[width=\textwidth]{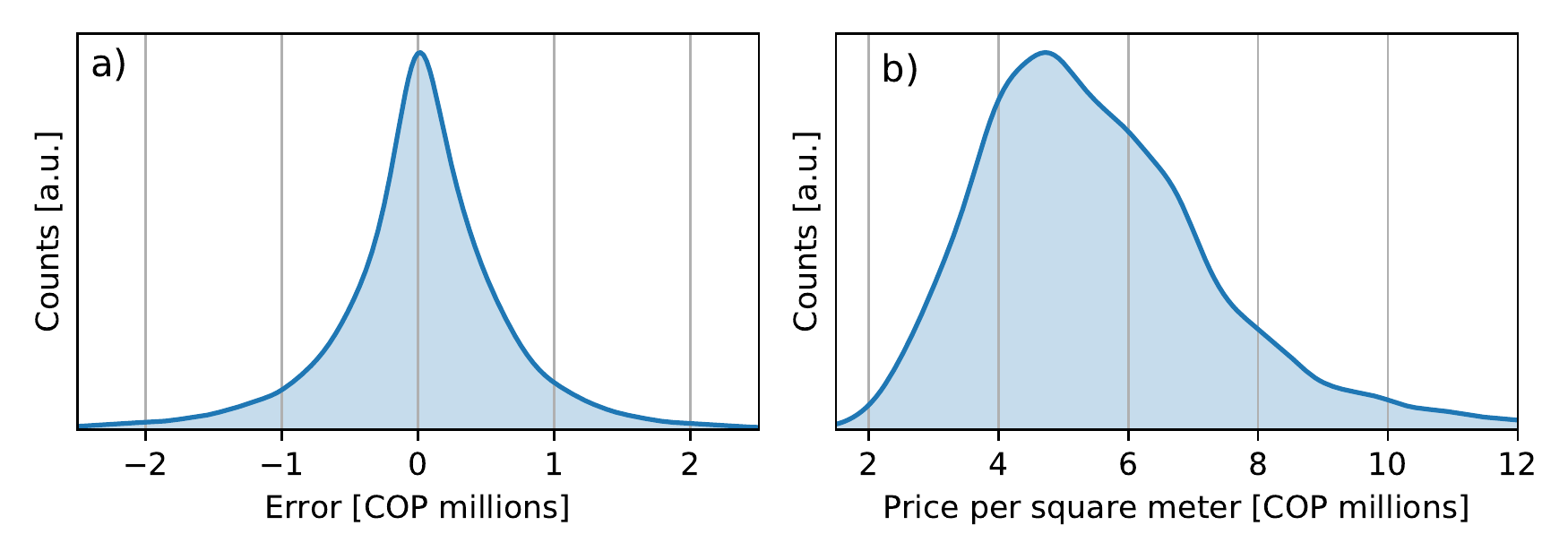}
    \caption{a) shows the error distribution of test set, and b) shows the price distribution per square meter of the test set.}
    \label{fig:error_dist_test}
\end{figure}

After the training of the XGBoost regressor, the mean error was measured on the test set, yielding an average error of approximately 0.5 million COP per square meter. This error corresponds to 8.9\% of the average price per square meter. The magnitude of the error has to be compared with the natural price variance offered by the market. In order to make such comparison, we group flats by neighbourhood, measure the price standard deviation and average the price standard deviation of all neighbourhoods. This average is approximately 1 million COP per square meter, corresponding to 18.97\% of the average price per square meter. With this characteristic scale of variance in the market, it is possible to understand how accurate our model is. A distribution plot of the error is shown in \cref{fig:error_dist_test}a), which is slightly skewed.

\begin{figure}[!ht]
    \centering
    \includegraphics[width=0.8\textwidth]{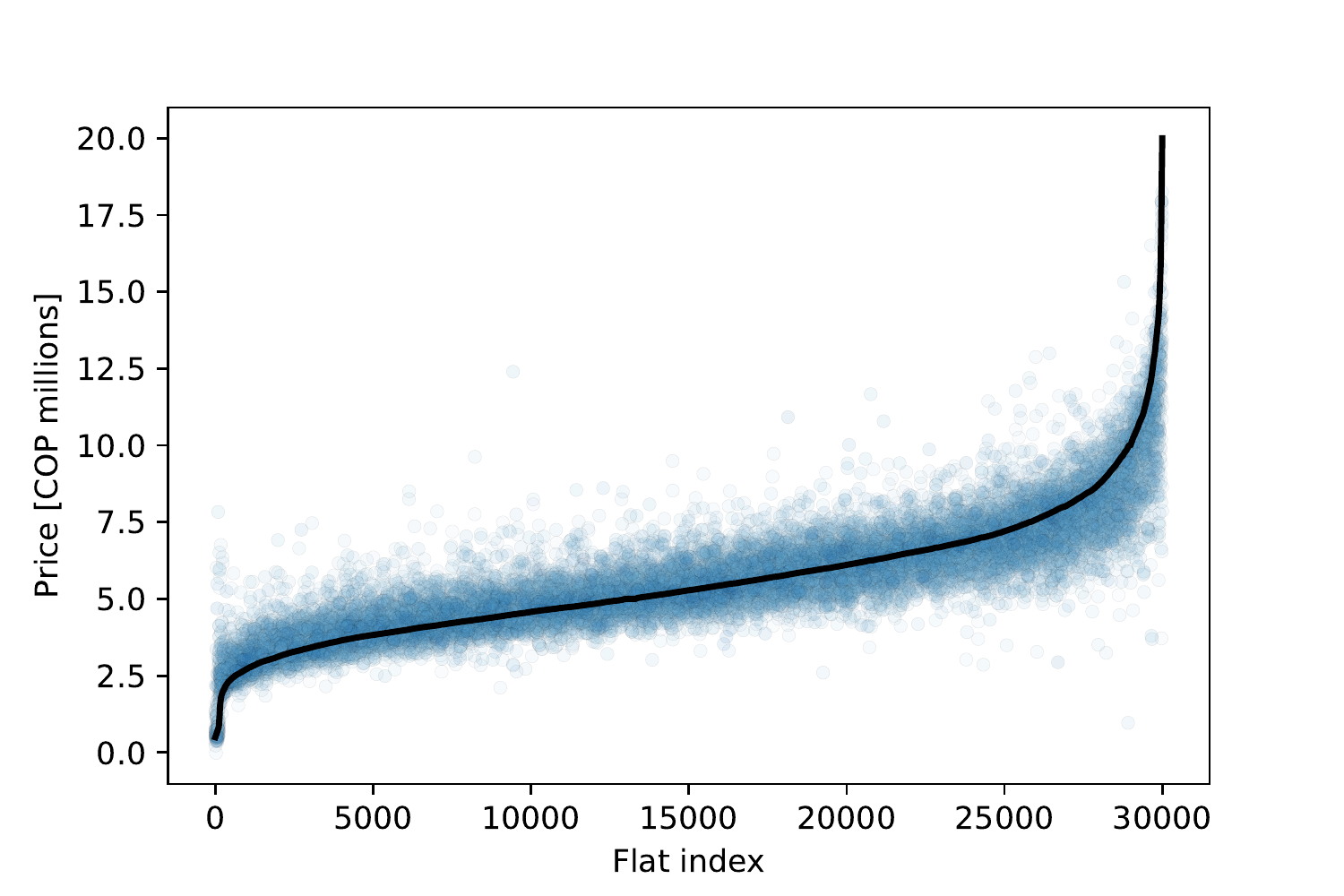}
    \caption{Flats in the test set are ordered from the cheapest to the most expensive one. Values refer to price per square meter. The black line shows the actual price, and the points show the predicted price for each of the flats.}
    \label{fig:price_dist}
\end{figure}

However, the asymmetry in the error can be clearly seen when examining the error for each price sector. In fact, the price distribution contains few flats at the tails, as it can be seen in \cref{fig:error_dist_test}b). As the sample is imbalanced, in the sense that there are more flats in the 3 to 8 million COP per square meter region, it is expected that the model overestimates the price of cheap flats, whereas it underestimates the price of expensive flats, as it is shown in~\cref{fig:price_dist}. However, it is seen that the estimated price closely wraps around the true price for all the regions of prices, even though the dispersion is slightly increased and skewed at the tails.

\begin{figure}[!ht]
    \centering
    \includegraphics[width=0.8\textwidth]{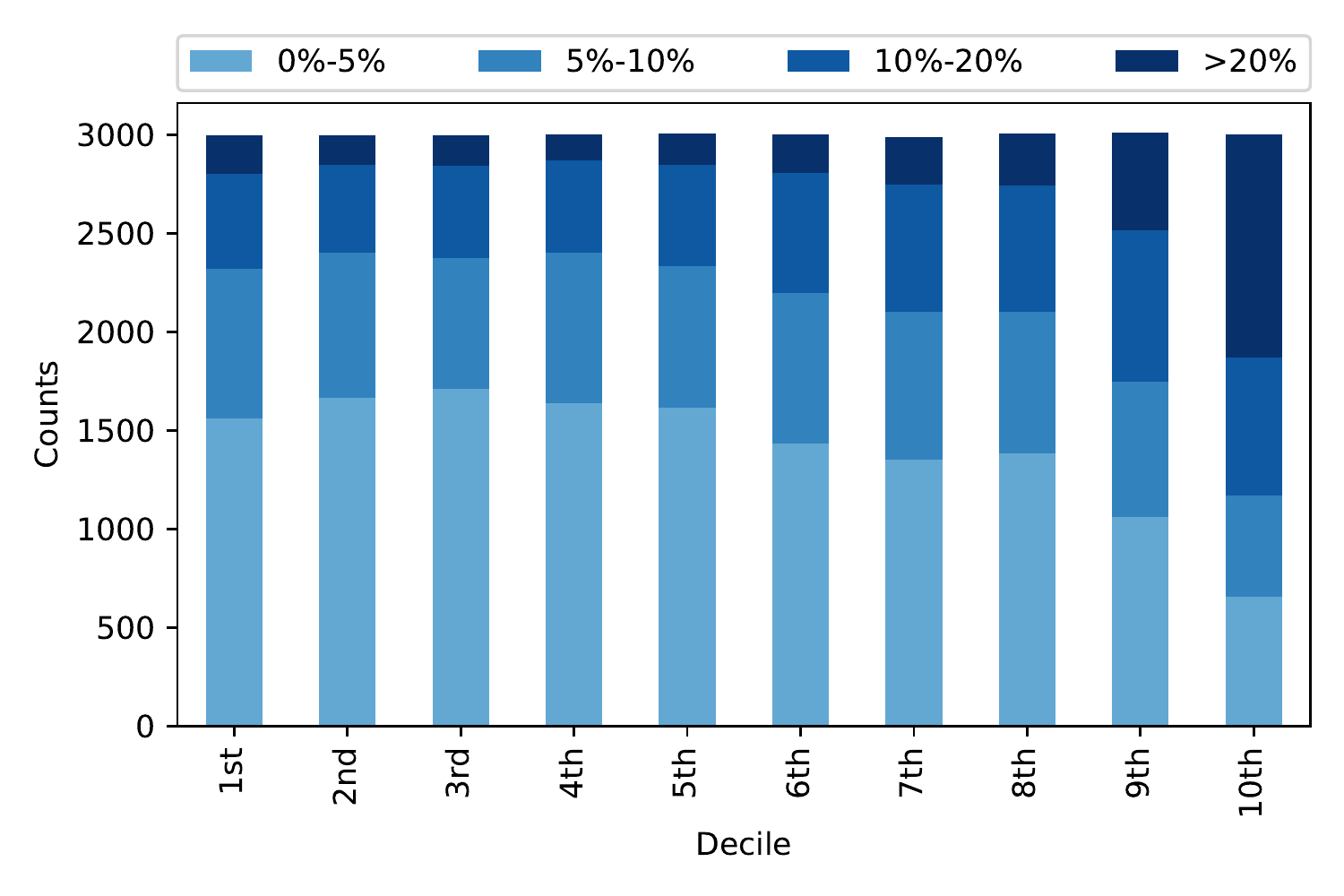}
    \caption{Error percentage distribution of flats in the test set for every price per square meter decile.}
    \label{fig:errordecile}
\end{figure}

A deeper examination of the error can be done partitioning the test set in price deciles. \Cref{fig:errordecile} shows the number of flats in the test set whose price estimation falls within a certain absolute error category for all the price deciles. The error categories are small (0\%-5\%), middle (5\%-10\%), large (10\%-20\%) and very large (greater than 20\%). It is seen that the error distribution of the model is robust up to the 8th decile, where the error starts to grow as flats become more expensive in the last two deciles. Since the model provides a range for the price, the true price of a flat will be found with high probability within the given range, even when the error distribution is more disperse as in the last two deciles.

\begin{figure}[!ht]
    \centering
    \includegraphics[width=0.8\textwidth]{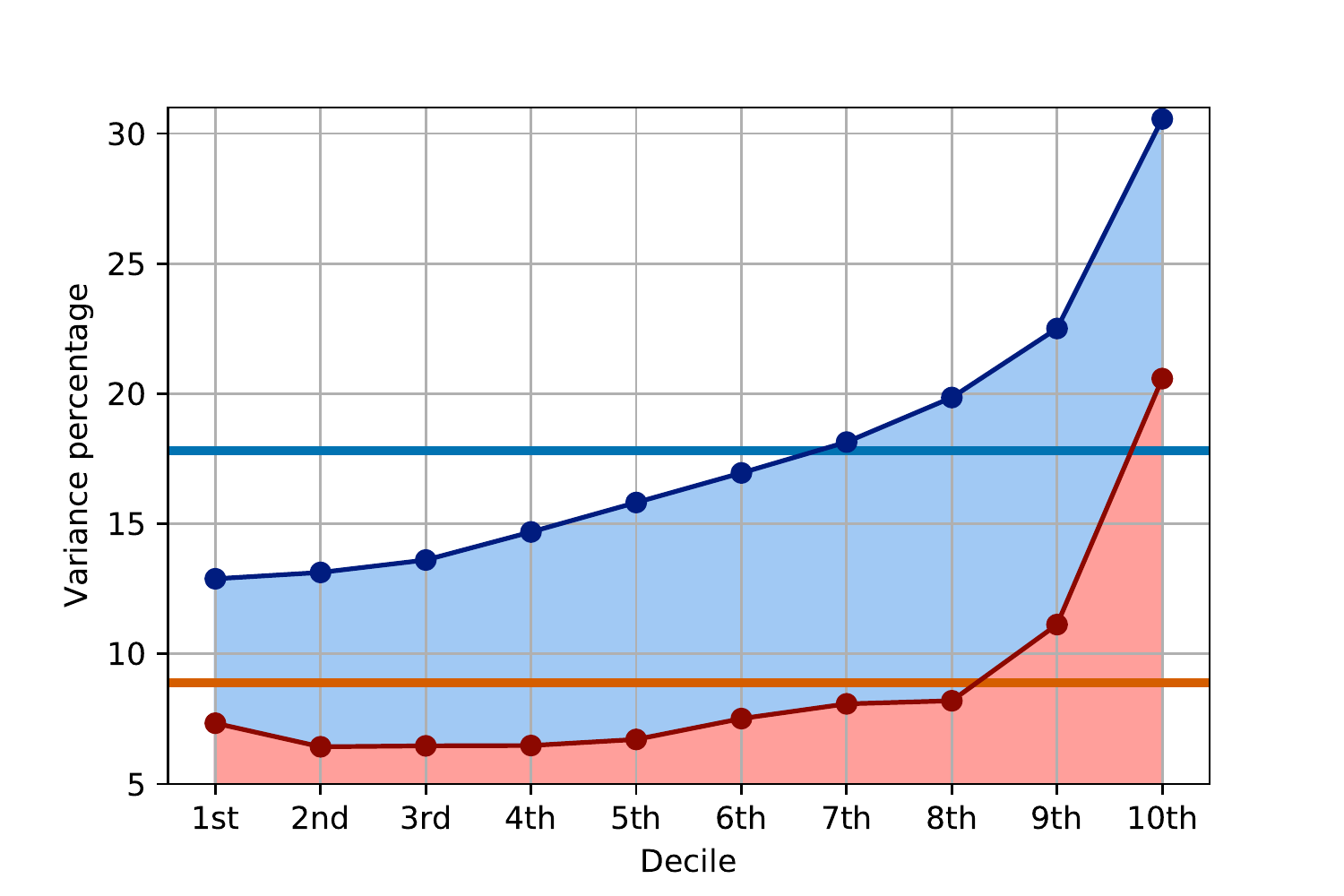}
    \caption{In blue, it is shown the mean-variance percentage with respect to the mean price of the flats in the test set for every decile. The horizontal blue line is the mean-variance percentage for all the flats in the database. Similarly, data in red refers to mean absolute error, with the same definitions as for the data in blue.}
    \label{fig:variance_percentage}
\end{figure}

Again, the error distribution throughout the deciles can only be understood when it is compared with the natural price dispersion scales found in the market. Previously, an average of the price standard deviation of all neighbourhoods was taken, but a more precise way of estimating the standard deviation consists in querying the property database for each flat in the test set, retrieving a set of its 20 closest flats. A standard deviation of the price is measured on this set and is finally averaged in the groups defined by the deciles. \Cref{fig:variance_percentage} shows the expected price variance in the submarkets defined by mathematical neighbourhoods of 20 flats centred at flats corresponding to each decile. In order to make a comparison between the natural market scale and the errors achieved by our model, the mean absolute error is also plotted. To state this formally, each blue point in \cref{fig:variance_percentage} is computed as
\begin{align}
    \frac{1}{P N_d}\sum_{n_d} \overline{\mathcal{N}}_{n_d}\times 100\%, && \overline{\mathcal{N}}_{n_d} = \frac{1}{20}\sum_{k=1}^{20} y_k(n_d),
\end{align}
where $N_d$ is the number of flats in the decile $d$, $n_d$ indexes those flats, $\overline{\mathcal{N}}_{n_d}$ is the average price of the flats in the mathematical neighbourhood of the flat indexed by $n_d$, $P$ is the average price of the flats in the test set, and $y_k(n_d)$ is the price of the $k$-th flat from the mathematical neighbourhood of the flat indexed by $n_d$.

Therefore, it is seen that the mean absolute error of our model is well below the natural price variance scales of the market for every decile. It is worth mentioning that the mean-variance percentage retrieving mathematical neighbourhoods of 20 flats for every flat in the test set is 17.82\%, which is slightly smaller than the 18.97\% found by retrieving all the flats in the physical named neighbourhood of every flat in the test set.

\begin{figure}[!ht]
    \centering
    \includegraphics[width=0.65\textwidth]{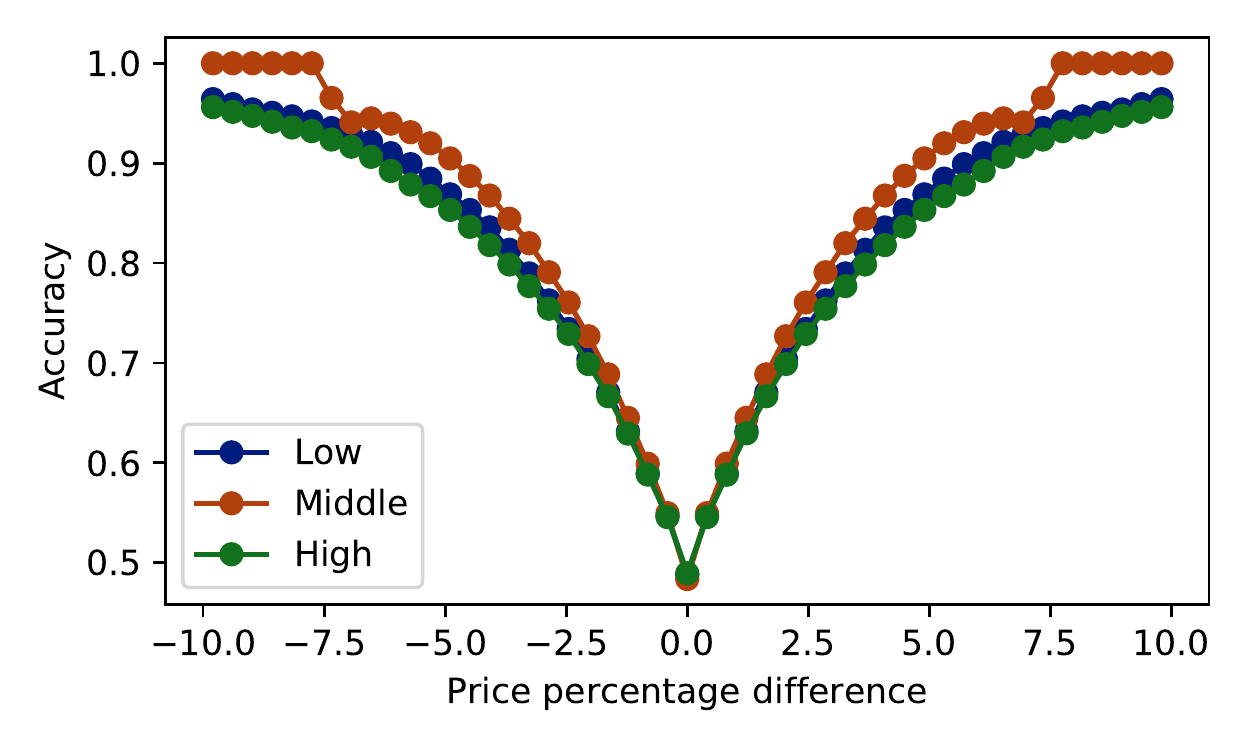}
    \caption{Accuracy of getting $\text{sgn}(\hat{y}_a - \hat{y}_b) = \text{sgn}(y_a - y_b)$ for a given pair of flats $a$ and $b$ as a function of the price percentage difference of that pair of flats.}
    \label{fig:pricepercdiff}
\end{figure}

A final test of how the model performs can be done by measuring how it preserves differences between the prices of flats. Let $y_a$, and $y_b$ be the prices per square meter of two flats $a$ and $b$ in the test set, respectively. Let $\hat{y}_a$ and $\hat{y}_b$ be the corresponding price estimations by the model. We want to check how the differences $y_a-y_b \stackrel{?}{=}\hat{y}_a-\hat{y}_b$ are preserved by the model. In order to do that, we partition the flats in the test set in three groups of low, middle and high prices per square meter, each one containing 10000 flats. For each group, we calculate the accuracy in predicting that if $y_a-y_b$ is positive or negative, so is $\hat{y}_a-\hat{y}_b$. These accuracies are shown in \cref{fig:pricepercdiff} as a function of the price percentage difference, which is defined as $(y_a-y_b)/[2(y_a+y_b)]$.

The same error structure is inherited to the socioeconomic stratum, which takes six values (from 1 to 6). However, there are only above 100 stratum 1 flats because that market does not usually use online platforms for advertising. Excluding those flats, the error remains from 6\% to 8\% in strata 2, 3, 4 and 5, while it grows to 11.5\% for stratum 6. This growth comes from the fact that the higher the stratum, the higher the price per square meter, and as it was seen in \cref{fig:errordecile,fig:variance_percentage}, the most expensive flats express the higher errors of the model.

Finally, we look at the error distribution of the price estimations in the test set as a function of the density of flats per unit area. Maps in \cref{fig:flatsinbogota} show a geographical concentration of the flats towards the northeast part of the city, whereas the density decreases in the west and south. By dividing the flat population in deciles by density, we are able to look at the error distribution as a function of density, which is shown in \cref{fig:densityerrdist}. Interestingly, error distribution is more uniform with respect to density changes.
\begin{figure}[!ht]
    \centering
    \includegraphics[width=0.8\textwidth]{Error_density_decile.pdf}
    \caption{Error percentage distribution of flats in the test set for every density decile.}
    \label{fig:densityerrdist}
\end{figure}

\begin{figure}
    \centering
    \includegraphics[width=0.8\textwidth]{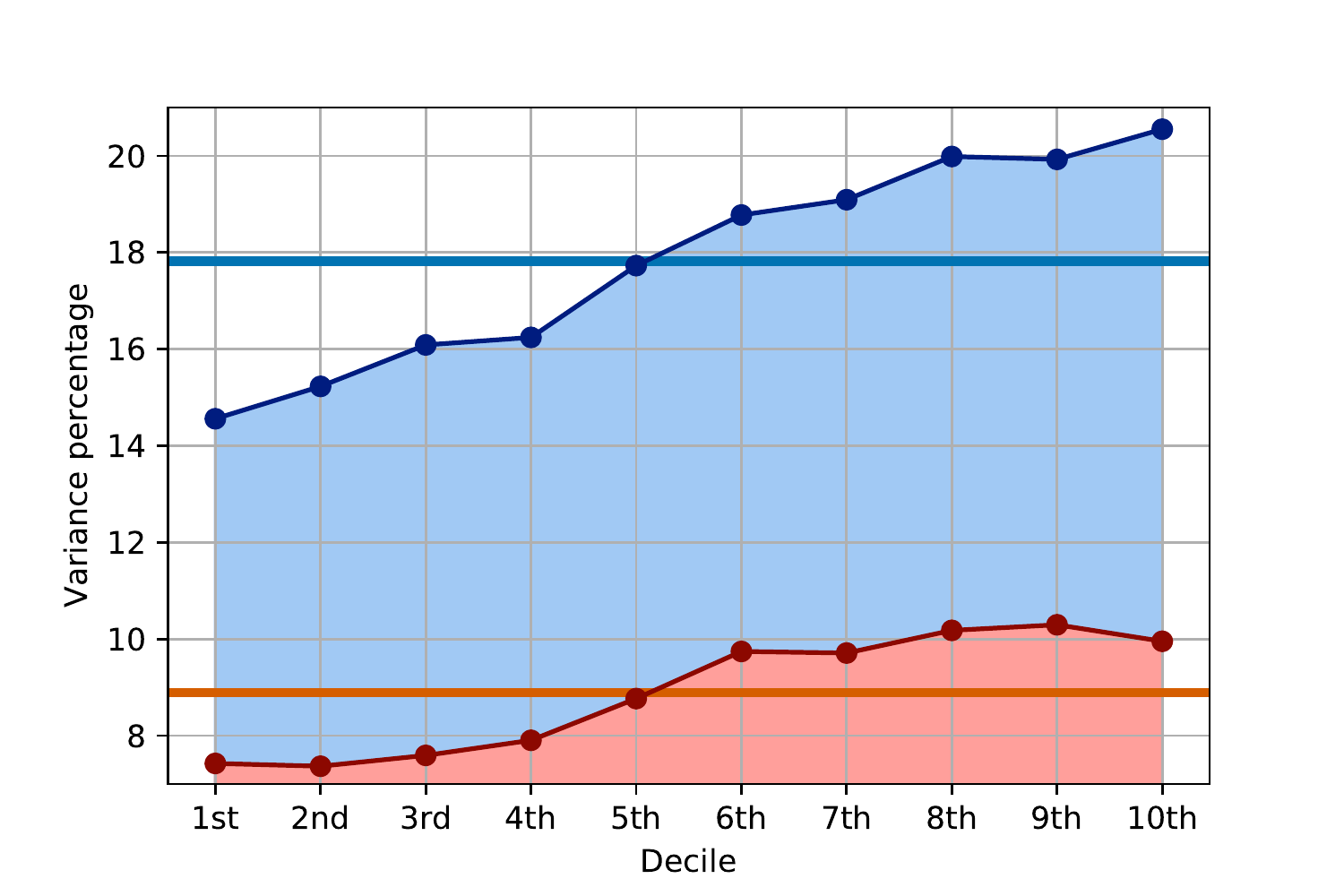}
    \caption{Same as \cref{fig:variance_percentage} but for density deciles.}
    \label{fig:vardensperc}
\end{figure}

Particularly, the model performs better at low flat densities, while the performance decreases for high flat densities. This can be explained because flats belonging to areas with high flat densities are at the northeast part of the city, as \cref{fig:flatsinbogota} suggests, which correspond to the highest prices per square meter. By our previous analysis, these high price zones have a high price variance. In fact, the error distribution has to be compared with the natural price variance scales of the market, this time segmented by density. 
\Cref{fig:vardensperc} shows this behaviour, where the densest zones correspond to high price variances indeed.

\section{Conclusions\label{sec:conclusions}}

Conventional real estate appraisal methods rely on the appraisal professional abilities to collect and interpret market data. It is not uncommon that two appraisers independently report different values for the same real estate property because they collected different sets of data or they interpreted data differently. This subjectivity inherent to conventional real estate appraisal methods gives room for unfairness in the market, as a stakeholder might take decisions based on flawed judgements made by an appraiser. Also, the appraising process is costly, as one needs to devote the time of an appraiser to collect and interpret market data. In the context of cities, where there are large data sets of real estate properties, conventional real estate appraisal methods can be avoided by replacing them (at least partially, as human intervention is always needed to supervise that results make sense) with state-of-the-art techniques for regression problems, such as estimating the price of a property.

The adoption of models based on state-of-the-art machine learning techniques for building appraising methods in the real estate market is of considerable importance, particularly for cities in developing countries, where real estate market regulations tend to be weak. To avoid speculation and subjectivity we proposed and developed an appraising method based on XGBoost that is robust, fair, and driven by a large property data set. Such data sets can be obtained from online real estate platforms, which are normally available in every city. We tested our proposal with almost 200,000 records of flats in Bogotá - Colombia collected from 2016 to 2020 including information about the flats' features and their neighbourhood features.

We analysed the robustness of our model by testing how well it performs on several segments. In general, the model has a mean error of 8.9\% when predicting price per square meter. Our model seems to perform worse for the most expensive real estate properties, as the error percentages get larger. However, we noted that the inherent variability of the price also increases for the most expensive real estate properties in our dataset. When comparing the increase of our model error to the increase of price variability for the segment of the most expensive properties, we see that our model error is always below the price variability of that segment. Moreover, as there are more dense parts of the city, we tested how the model performed when estimating prices of properties on high and low housing density. One would think that the denser the neighbourhood of a property, the more data, and therefore, the more precise the model would be. However, this was shown to be wrong. Denser areas have properties whose price variability is larger than sparse areas. This fact is correlated with the increase in price variability for expensive properties, as these are usually found in dense areas. The error of our model also shows an increasing trend (but always below the inherent price variability) as density gets larger. As the research has demonstrated, the model performs well for every segment of price and housing density that was considered in our study.

We urge local governments in Colombia and in cities from developing countries to promote robust real estate pricing models that are based on large data sets available to governments, which improve appraisal accuracy, increase their fairness, and avoid price speculation.




\bibliographystyle{apacite}
\bibliography{sample}

\end{document}